\documentclass{article} 
\usepackage{arxiv_journal,times}

\usepackage{amssymb}
\usepackage{amsthm}

\usepackage{bbm}
\usepackage{hyperref}
\usepackage{url}
\usepackage{graphicx}
\usepackage{physics}
\usepackage{amsfonts}
\usepackage{wrapfig}
\usepackage{changepage}
\usepackage{color,xcolor}
\usepackage{graphicx}
\usepackage{caption}
\usepackage{subcaption}
\usepackage{cite}

\usepackage{booktabs}
 
\title{Scalable Neural Decoders for Practical Real-Time Quantum Error Correction}

\iclrfinalcopy
\author{Changwon Lee\thanks{These authors contributed equally.}
\\
Department of Statistics and Data Science\\
Yonsei University\\
Seoul, Republic of Korea \\
\texttt{changwonlee@yonsei.ac.kr} \\
\And
Tak Hur\footnotemark[1]
\\
Department of Statistics and Data Science\\
Yonsei University\\
Seoul, Republic of Korea \\
\texttt{takh0404@yonsei.ac.kr} \\
\And
Daniel K. Park \\
Department of Applied Statistics \\
Department of Statistics and Data Science \\
Department of Quantum Information \\
Yonsei University \\
Seoul, Republic of Korea \\
\texttt{dkd.park@yonsei.ac.kr} \\
}

\iclrfinalcopy
\begin{document}

\maketitle
\begin{abstract}
Real-time, scalable, and accurate decoding is a critical component for realizing a fault-tolerant quantum computer. 
While Transformer-based neural decoders such as \textit{AlphaQubit} have demonstrated high accuracy, the computational complexity of their core attention mechanism, which scales as $\mathcal{O}(d^4)$ with code distance $d$, results in decoding speeds insufficient for practical real-time applications. 
In this work, we introduce and evaluate a \textit{Mamba}-based decoder, a state-space model with $\mathcal{O}(d^2)$ complexity. 
In memory experiments using Sycamore hardware data, our Mamba decoder matches the performance of its Transformer-based counterpart, providing that its superior efficiency does not come at the cost of performance. 
Crucially, in simulated real-time scenarios that account for decoder-induced noise, the Mamba decoder significantly outperforms the Transformer, exhibiting a higher error threshold of $0.0104$ compared to $0.0097$.
These results demonstrate that Mamba decoders offer a compelling balance between speed and accuracy, making them a promising architecture for scalable, real-time quantum error correction.
\end{abstract}

\section*{Introduction}
\label{sec:introduction}
Quantum computing promises computational capabilities that surpass those of classical systems, yet realizing these advantages at scale requires effective strategies to mitigate noise and imperfections. Quantum error correction (QEC) addresses this challenge by encoding logical information across multiple physical qubits and performing repeated syndrome measurements to detect errors~\cite{lidar2013quantum}. 
A classical decoder processes these syndrome outcomes to infer the most likely error and instructs appropriate corrections, thereby preserving the logical qubit state. 
Crucially, decoding must be both accurate and fast: while the decoder computes a correction, the physical qubits remain exposed to noise, allowing additional errors to accumulate. 
Therefore, operating within the hardware's error correction cycle time is a minimum requirement for decoders to avoid accumulating a backlog of syndrome data~\cite{terhal2015quantum}; faster operation is vital to reduce latency and suppress errors during fault-tolerant computation.
As quantum processors scale to larger codes and more logical qubits, achieving low-latency, high-throughput decoding becomes increasingly critical.

A variety of decoding algorithms have been developed to meet these requirements. 
Traditional matching-based decoders---such as minimum-weight perfect matching (MWPM)~\cite{higgott2022pymatching}, belief matching~\cite{PhysRevX.13.031007}, and correlated matching~\cite{fowler2013optimal}---can correct errors quickly, but do not fully capture complex correlations in the syndrome data. 
Meanwhile, more exhaustive approaches like tensor-network (TN) decoders~\cite{ferris2014tensor} can provide highly accurate solutions but scale poorly with code distance, unless strong bond-dimension truncations are applied, which in turn degrade accuracy.
To overcome these trade-offs, machine learning techniques have recently emerged as a promising alternative for decoding~\cite{overwater2022neural, wang2022multidimensional, liu2019neural, lange2025data, baireuther2019neural, baireuther2018machine, chamberland2018deep, cao2023qecgpt, egorov2023end, fitzek2020deep, gicev2023scalable, krastanov2017deep, maskara2019advantages, ni2020neural, sweke2020reinforcement, torlai2017neural, varbanov2025neural, varsamopoulos2019comparing, varsamopoulos2020decoding, wagner2020symmetries, zhang2023scalable, varsamopoulos2017decoding}.

Among these, \emph{AlphaQubit}~\cite{bausch2024learning} represents a major milestone, demonstrating state-of-the-art decoding performance in memory experiments. Leveraging a recurrent Transformer-based architecture, AlphaQubit learns to predict logical errors from syndrome data, surpassing existing decoders in both accuracy and adaptability. 
Through a two-stage training process---pretraining on synthetic noise models followed by fine-tuning on real experimental data---the model adapts to complex, hardware-specific noise sources, such as cross-talk, leakage, and soft readouts. 
This ability to directly process soft information and account for non-standard error channels like leakage provides a distinct advantage over traditional decoders, which are often restricted to binary syndrome outcomes~\cite{pattison2107improved}.

In memory experiments using real hardware data from the Sycamore processor for distance-3 and distance-5 surface codes, AlphaQubit outperformed all other decoders, including the highly accurate TN decoder. 
For larger codes up to distance 11, which are inaccessible on current devices, evaluations were performed on simulated data. 
In these simulations, AlphaQubit continued to outperform correlated matching decoders. 
This success underscores the power of data-driven adaptation and demonstrates how machine learning can push QEC decoding beyond the limits of human-designed algorithms.

However, AlphaQubit also exposes a fundamental limitation: the computational burden of its Transformer-based architecture~\cite{vaswani2017attention}. 
The self-attention mechanism at the core of the Transformer scales as \(\mathcal{O}(n^2)\) with input length, leading to an overall \(\mathcal{O}(d^4)\) for surface codes, as the syndrom bitstring length scales with \(d^2 - 1\). 
While AlphaQubit performs well in memory experiments, where decoding is deferred until the end, its $\mathcal{O}(d^{4})$ complexity makes it impractically slow for real-time error correction. Superconducting quantum processors require immediate feedback corrections on a microsecond-scale threshold. For distance-5 codes, AlphaQubit's latency is already an order of magnitude slower than this requirement, and this speed penalty only worsens as the code distance increases.
This bottleneck highlights the need for an alternative architecture that achieves comparable accuracy while drastically improving inference speed.

\begin{figure}[t]
    \centering 
    \includegraphics[width=0.9\textwidth]{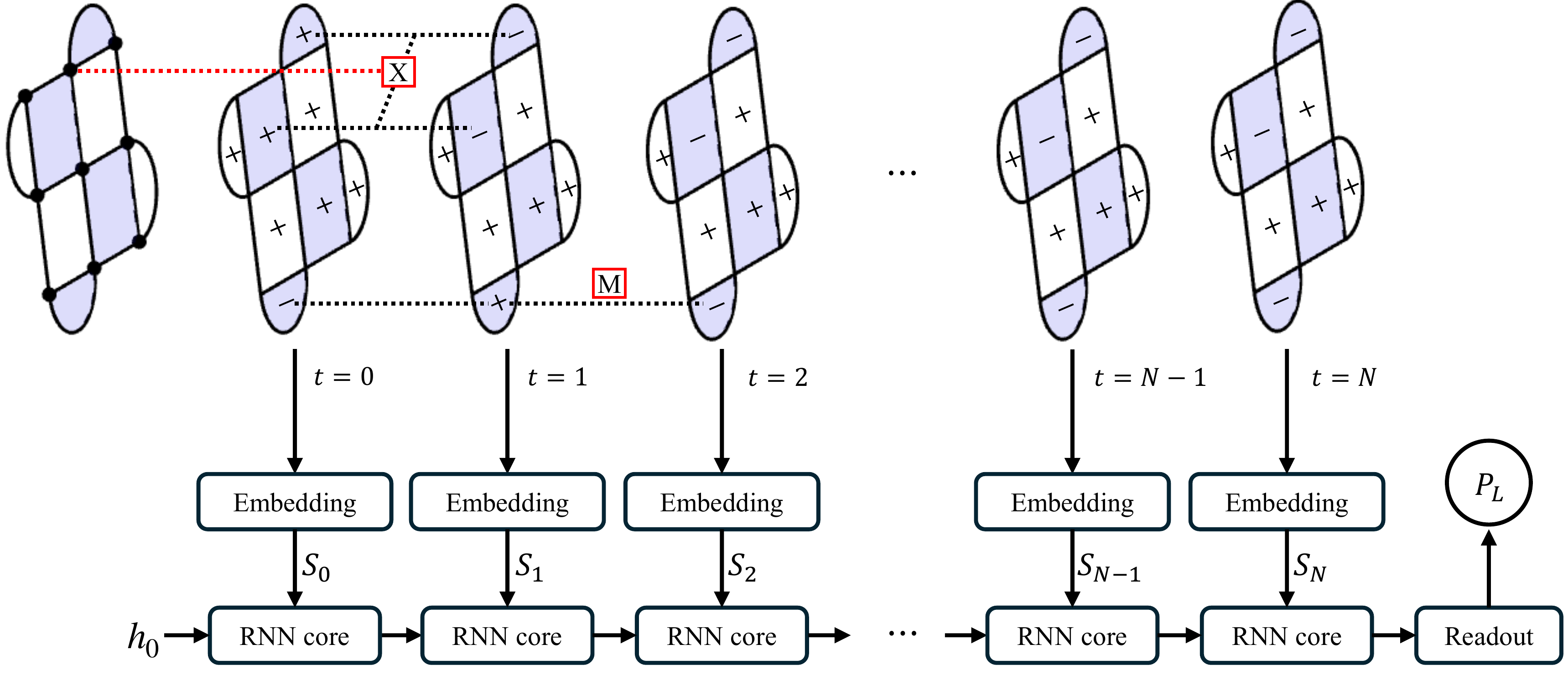}
  \caption{The overall recurrent architecture of the neural decoder. At each cycle $n$, stabilizer measurement embeddings $S_n$ are processed by an RNN core, updating the decoder's hidden state from $h_n$ to $h_{n+1}$. The final state is passed to a readout network to predict $P_{L}$, the probability of a logical error. Further details on the architecture and notations are provided in the Methods section. 
  }
    \label{fig:fig1}
\end{figure}

In this work, we introduce and analyze the \emph{Mamba decoder}, a novel neural decoder architecture designed to address the computational bottlenecks of Transformer-based models like AlphaQubit. Our decoder is built upon Mamba~\cite{gu2023mamba}, a state-space model that replaces the explicit attention mechanism with selective state updates.
By maintaining a continuous representation of historical context in the decoder's hidden states, Mamba avoids full pairwise interactions and reduces computational complexity from quadratic to linear, while still capturing essential correlations within the syndrome data.
This architecture is particularly well-suited for QEC decoding, which requires efficient processing of long syndrome sequences. 
This mechanism makes our Mamba-based decoder an effective solution that balances the need to model complex error chains with the stringent low-latency requirements of real-time decoding.
The schematic of the overall architecture is provided in Figure~\ref{fig:fig1}.

Our investigation proceeds in two parts. First, we validate our Mamba decoder by applying it to the Sycamore memory experiments, a key benchmark for quantum error correction. Using the publicly available hardware dataset for distance-3 and distance-5 codes, we test our decoder's performance on real-world experimental data. In this setting, our Mamba decoder's performance matches that of the Transformer-based architecture, demonstrating that its superior computational efficiency does not come at the cost of performance.
Second, we shift our focus to the more practical real-time decoding scenario. 
We simulate this by introducing decoder-induced noise at each correction round, with the noise strength scaling according to the decoder’s computational complexity—$\mathcal{O}(d^2)$ for Mamba and $\mathcal{O}(d^4)$ for the Transformer, a choice justified by wall-clock time comparisons. 
In these simulations, using the SI1000 noise model, the Mamba decoder outperforms the Transformer for distances 3 and 5. 
Furthermore, by analyzing the error threshold under these real-time conditions, we find that the Mamba decoder exhibits a significantly higher threshold (0.0104) compared to the Transformer decoder (0.0097).
These results strongly suggest that Mamba-based architectures are a compelling choice for scalable, real-time decoding, a crucial component for the realization of a fault-tolerant quantum computer.

\section*{Results}
\subsection*{Memory Experiments}

To evaluate the performance of our Mamba-based decoder, we benchmark it against Transformer-, matching-, and TN–based decoders using the Sycamore memory experiment dataset from AlphaQubit~\cite{bausch2024learning}.
This publicly available dataset, provided by Google Quantum AI~\cite{google2023suppressing}, was obtained on a 72-qubit superconducting processor comprising four distance-3 surface-code blocks and a single distance-5 block.
Both $X$- and $Z$-basis memory experiments were conducted for up to 25 cycles of error correction, with 50,000 experiments performed for each odd-numbered cycle count $n \in \{1, 3, ..., 25\}$. This setup is designed to trace the logical qubit's fidelity as it decays over an increasing number of correction cycles.

Decoder performance is quantified by the logical error per round (LER), which indicates the probability that the decoder will fail per error correction cycle. 
To estimate the LER, we perform a linear regression on the logarithm of the measured fidelities as a function of the number of error correction cycles $n$, using the model $\log{F(n)} = \log{F_{0}} + n \log{(1 - 2 \epsilon)}$. 
Here, $\epsilon$ corresponds to the logical error per round.  

\begin{figure}[t]
  \centering
    \begin{subfigure}[b]{0.66\textwidth}
        \includegraphics[width=1\linewidth]{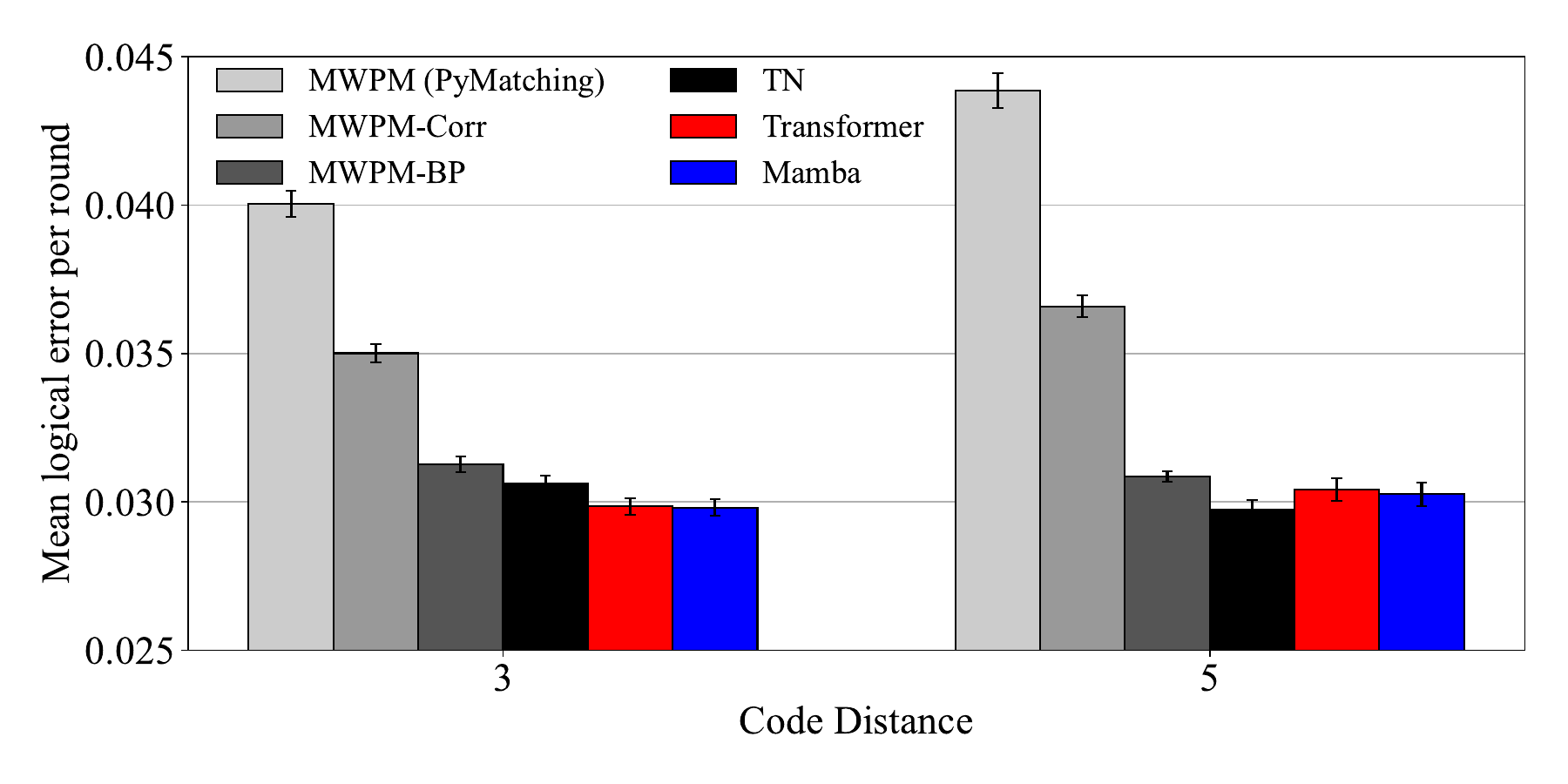}
        \caption{}
        \label{fig:fig2_1}
    \end{subfigure}% 
    \begin{subfigure}[b]{0.33\textwidth}
        \includegraphics[width=1\linewidth]{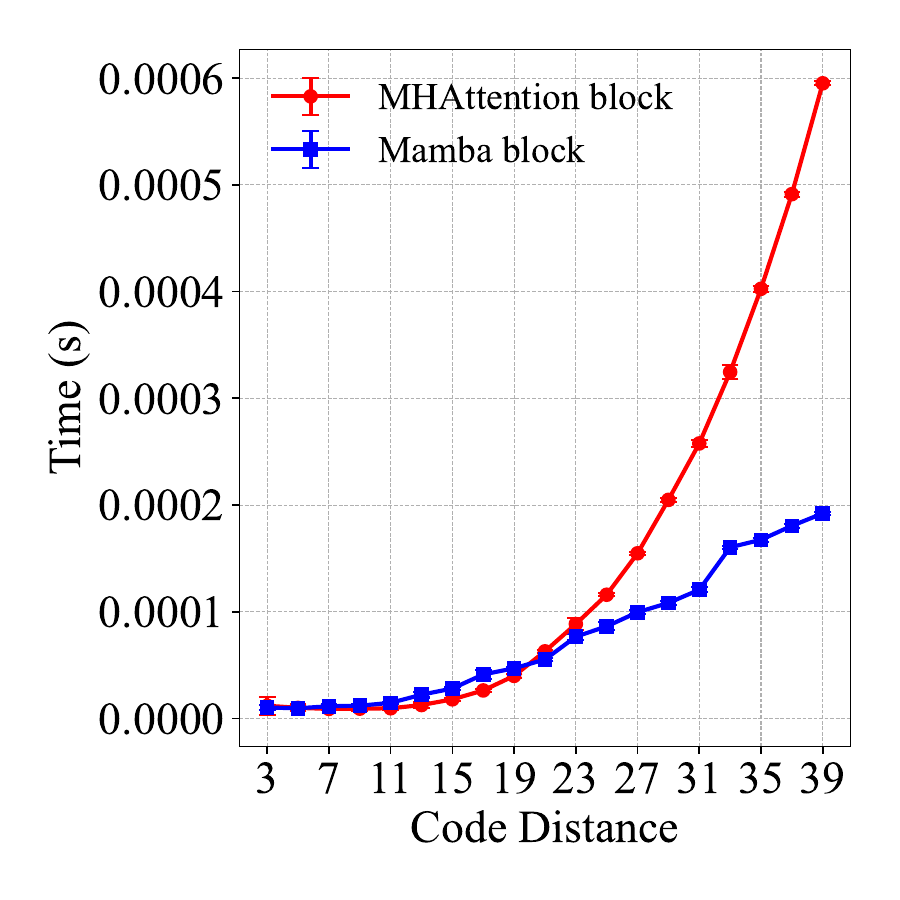}
        \caption{}
        \label{fig:fig2_2}
    \end{subfigure}% 
  \caption{(a) Logical error per round on the Sycamore dataset for various decoders at code distances 3 and 5. (b) A comparison of inference time for a Mamba block versus a Multi-Head Attention (MHA) block as code distance increases, measured on a local RTX 4090 GPU. 
  }
  \label{fig:fig2}
\end{figure}

Our training pipeline consists of two stages: pretraining and fine-tuning. 
In the pretraining stage, we generate up to 100 million synthetic samples using a detector error model (DEM) whose weights are derived from a Pauli noise model~\cite{bausch2024learning}. 
The parameters of this noise model are calibrated using cross-entropy benchmarking (XEB) data to approximate the noise characteristics of the target quantum hardware. 
In the fine-tuning stage, we adapt the pretrained decoder using half of the Sycamore experimental dataset, while holding out the remaining half for evaluation. 
The fully trained decoder is subsequently evaluated on the held-out set of 25,000 samples. 
The results are shown in Figure~\ref{fig:fig2_1}.

Our Mamba decoder achieves LERs of approximately \(2.98\times10^{-2}\) at distance 3 and \(3.03\times10^{-2}\) at distance 5, matching the performance of Transformer decoders. 
However, at distance 5 both the transformer and Mamba decoder underperform compared to the tensor network decoder~\cite{bravyi2014efficient}. 
We attribute this performance gap to the size of the pretraining dataset. 
While AlphaQubit was pretrained on up to 1 billion samples, our models were pretrained on 100 million samples due to computational constraints. 
We project that with a larger pretraining dataset, our neural decoders could match or exceed the performance of the tensor network decoder.

It is important to note that while both decoders deliver comparable logical error rates, their computational scaling differs dramatically. 
Figure~\ref{fig:fig2_2} shows wall-clock inference time on a local RTX 4090 GPU for a Mamba block versus a Multi-Head Attention (MHA) block used in the Transformer decoder as code distance increases. 
The Transformer decoder's inference cost scales as $\mathcal{O}(d^{4})$, a consequence of its core self-attention mechanism.
Because the number of stabilizer syndromes is proportional to $d^2$ for surface codes, and self-attention requires quadratic, pairwise comparisons between all syndromes, the computational cost grows as $d^4$. 
This leads to prohibitive latency at larger code distances. 
In contrast, our Mamba decoder, as a state-space model, avoids these expensive pairwise computations~\cite{gu2021efficiently}. 
It processes the syndrome sequence linearly by updating a compressed hidden state at each step, resulting in an overall complexity of $\mathcal{O}(d^{2})$.
This provides a much more favorable speed-accuracy trade-off, making it a more viable candidate for real-time quantum error correction.

\subsection*{Real Time Decoding with Decoding Noise}
\begin{figure}[t]
    \centering
    \includegraphics[width=0.9\textwidth]{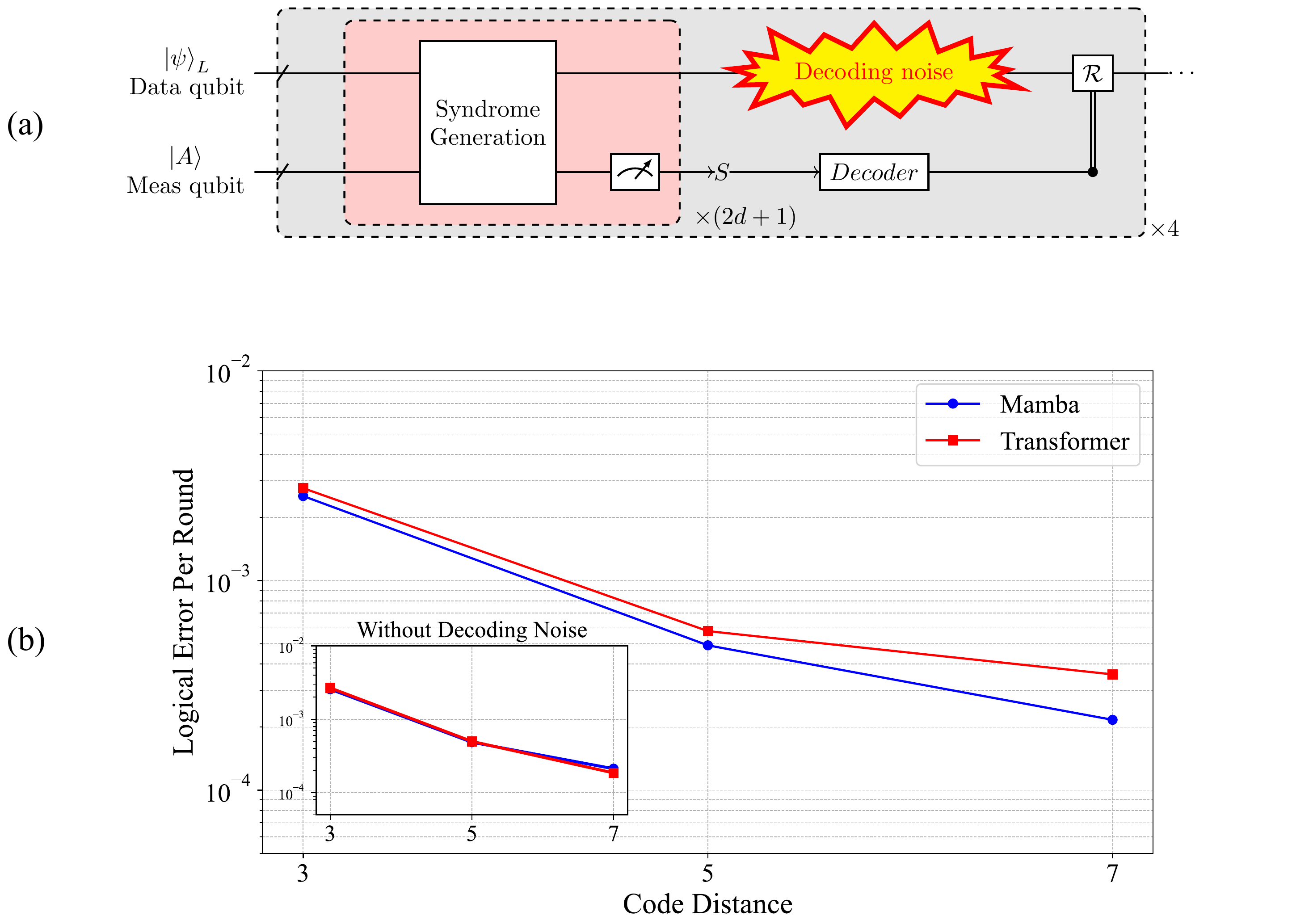}
    \caption{(a) Experimental scheme for real-time decoding simulation. The evaluation runs for $8d+4$ cycles, structured as four repetitions of a $2d+1$ cycle block. After each block, decoding noise is injected to simulate the impact of decoder latency before a correction $\mathcal{R}$ is applied. (b) Comparison of Mamba and Transformer decoders in terms of Logical Error per Round (LER) under real-time decoding scenarios, evaluated at code distances of 3, 5, and 7. Performance is shown with decoder-induced noise, whose strength is scaled according to each decoder's computational complexity. For reference, LERs obtained without decoder-induced noise are shown in the inset.}
    \label{fig:fig3}
\end{figure}
In a practical real-time QEC, a decoder's latency is a critical factor. 
A slow decoder can fall behind the hardware's clock cycle, allowing uncorrected physical errors to accumulate through subsequent cycles~\cite{terhal2015quantum}. 
This latency-dependent phenomenon, which we refer to as decoder-induced noise, is a significant challenge for maintaining logical fidelity during long computations.

We simulated real-time decoding for surface codes of distance 3, 5, and 7, using both the Mamba and Transformer decoders. 
The underlying physical errors were simulated using the Superconducting-Inspired Circuit Depolarizing Noise (SI1000) model~\cite{gidney2021fault} using Stim~\cite{gidney2021stim}.
In this model, we applied depolarizing errors with strengths of $p/10$, $p$, $2p$, and $5p$ to single-qubit gates, two-qubit gates, resets, and measurements, respectively, with a base physical error rate of $p=0.002$. 
Mirroring the memory experiments, each model was trained on sequences of $2d + 1$ QEC cycles. 
However, to evaluate performance in a prolonged real-time scenario, the logical error rate (LER) was computed over a much longer duration of $8d + 4$ cycles. 
Crucially, to simulate the effect of decoder latency, decoder-induced noise was injected every $2d + 1$ cycles throughout the evaluation period. 
This setup results in four distinct injections of decoder-induced noise, allowing us to measure the cumulative impact of latency on logical performance. 
The experimental schematic is shown in Figure~\ref{fig:fig3}a.

The strength of the decoder-induced noise was carefully calibrated to reflect the differing computational complexities of the two architectures. 
For large code distances ($d > 20$), which are essential for practical QEC, the overall inference time is dominated by the most computationally intensive components: the Multi-Head Attention (MHA) block in the Transformer and the Mamba block in our model. 
As shown by our wall-clock measurements (Figure~\ref{fig:fig2_2}), the latency of these blocks scales as $\mathcal{O}(d^{4})$ and $\mathcal{O}(d^{2})$, respectively. 
We therefore modeled the decoder-induced noise as an additional depolarizing channel whose error probability, $p_{\mathrm{dec}}$, is proportional to these complexities. 
Specifically, we define $p_\mathrm{dec} = \alpha \cdot d^{4}$ for the Transformer and $p_\mathrm{dec} = \alpha \cdot d^{2}$ for Mamba, where $\alpha$ is a scaling constant.
To determine a physically relevant value for $\alpha$, we used the reported approximately $40 \mu s$ decoding time for the AlphaQubit architecture at $d=9$ as a benchmark. 
Based on this, we made a modest approximation by setting the induced error strength for the Transformer at this distance to be $25p$, five times stronger than the physical measurement error ($5p$). 
This choice yields a scaling constant of $\alpha = 7.623 \times 10^{-6}$, which was then applied consistently across all distances for both decoder models.

Figure~\ref{fig:fig3}b shows the LERs for both decoders at distances 3, 5, and 7, both with and without the simulated decoder-induced noise. 
In the absence of this induced noise, the Mamba and Transformer decoders exhibit nearly identical performance, underscoring their comparable accuracy on the underlying physical noise model. 
However, the introduction of decoder-induced noise reveals a stark divergence. 
The Transformer's prohibitive $\mathcal{O}(d^{4})$ latency leads to a severe accumulation of errors, causing a significant degradation in its LER as the code distance increases. 
In contrast, the Mamba decoder's more favorable $\mathcal{O}(d^{2})$ scaling results in a much smaller penalty. 
This resilience to latency-induced errors allows it to substantially outperform the Transformer, confirming its superior scalability and robustness for practical, real-time quantum error correction.

\subsection*{Analysis of the Error Threshold under Real Time Decoding} 

\begin{figure}[t]
  \centering
    \begin{subfigure}[b]{0.5\textwidth}
        \includegraphics[width=1\linewidth]{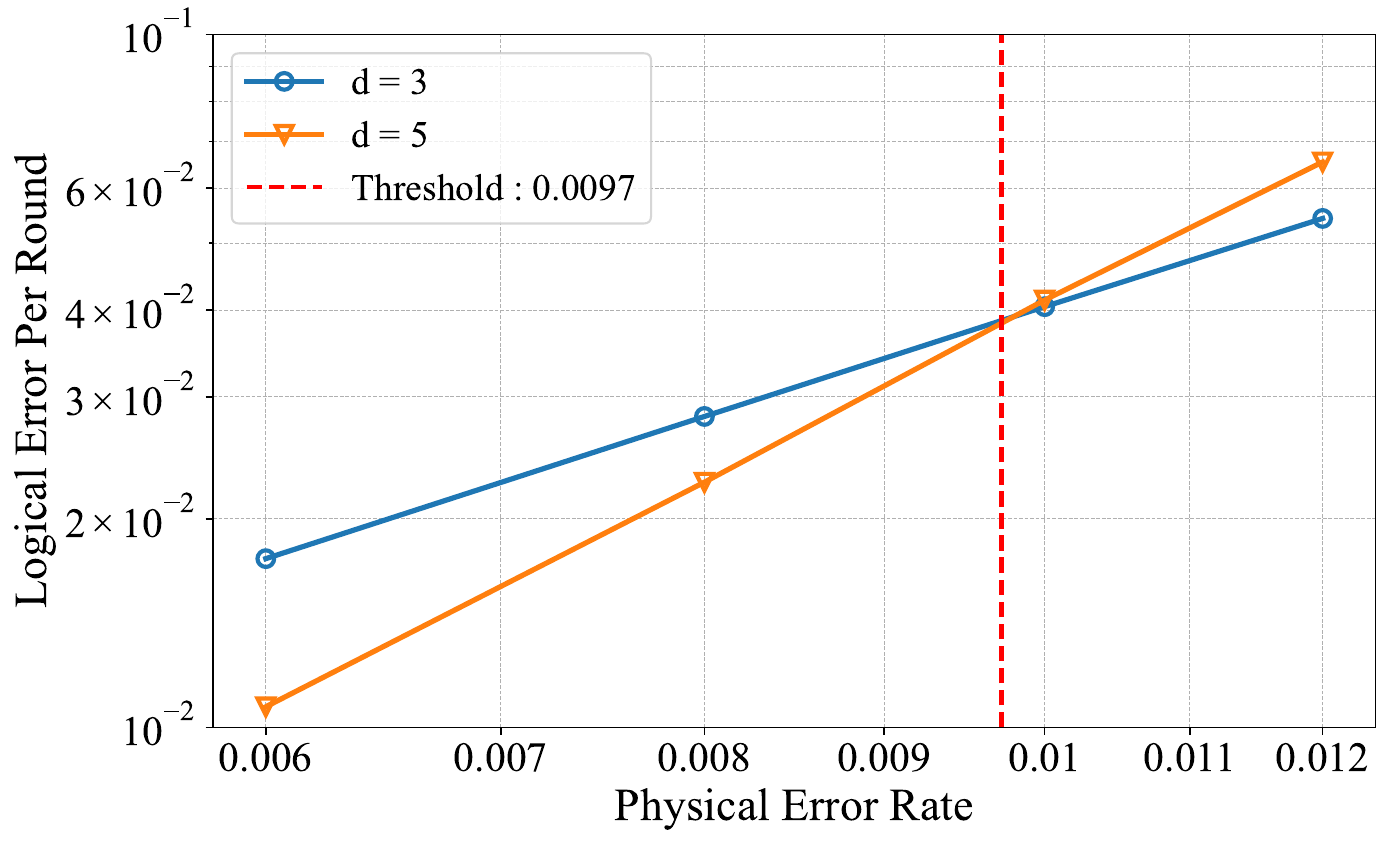}
        \caption{}
        \label{fig:fig4_1}
    \end{subfigure}% 
    \begin{subfigure}[b]{0.5\textwidth}
        \includegraphics[width=1\linewidth]{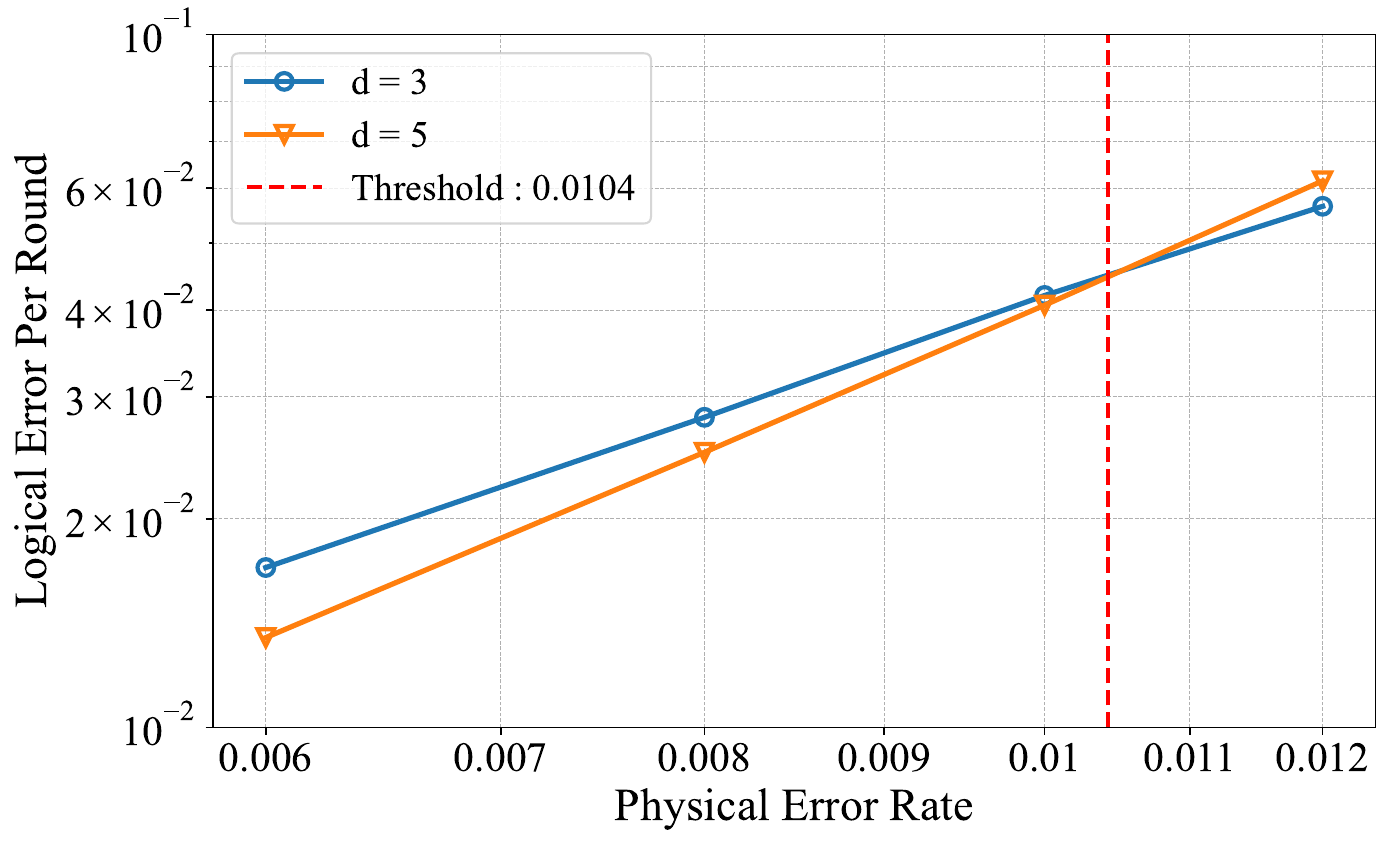}
        \caption{}
        \label{fig:fig4_2}
    \end{subfigure}% 
  \caption{Analysis of the effective error threshold under real-time decoding with decoder-induced noise. The plots show the Logical Error Per Round (LER) as a function of the Physical Error Rate ($p$) for code distances $d=3$ and $d=5$. The effective error threshold is the crossover point where the LER for $d=5$ (orange line) becomes higher than the LER for $d=3$ (blue line). (a) The Transformer-based decoder, showing a threshold of 0.0097. (b) The Mamba-based decoder, showing a higher threshold of 0.0104.
  }
  \label{fig:fig4}
\end{figure}

A critical metric for any quantum error correction scheme is the error threshold ($p_\text{th}$), defined as the maximum physical error rate which increasing the code distance ($d$) leads to an exponential suppression of the logical error rate. 
A decoder that exhibits a higher error threshold is fundamentally more robust, as it can achieve fault tolerance with noisier physical qubits, a significant advantage for near-term hardware. 
In a real-time setting, a decoder's computational latency introduces an additional noise channel—decoder-induced noise—which effectively lowers this threshold. 
Therefore, an ideal real-time decoder must not only be accurate but also fast enough to minimize this performance degradation.

We investigate the effective error threshold---that is, the threshold accounting for additional decoder-induced noise due to finite latency---of our Mamba decoder compared to the Transformer-based architecture under real-time decoding. 
This experiment is designed to determine the maximum physical error rate at which each decoder architecture can still provide effective error suppression. 
To ensure that each decoder achieves its optimal performance at each noise level, we employed a fine-tuning strategy. 
Starting with the baseline models pre-trained at a physical error rate of $p$=0.002, we fine-tuned separate model instances for each higher physical error rate: $p\in\{0.006,0.008,0.010,0.012\}$. 

The results, shown in Figure~\ref{fig:fig4} for code distances 3 and 5, demonstrate the substantial impact of computational efficiency on the effective error threshold. 
The defining characteristic of operating below the threshold is that a larger code distance yields a lower logical error rate ($P_L (d=5)< P_L (d=3)$). 
As shown in Figure~\ref{fig:fig4_1}, the LER curves for the Transformer architecture cross at a physical error rate of approximately $p_\text{th} \approx 0.0097$. 
In contrast, the Mamba decoder's curves (Figure~\ref{fig:fig4_2}) cross at a significantly higher physical error rate of $p_\text{th} \approx 0.0104$. 
This crossover point, found by comparing $d=3$ and $d=5$, represents the effective threshold for this specific comparison. 
However, as the code distance continues to increase, the unfavorable computational scaling of the Transformer decoder will cause this threshold to degrade rapidly. 
The Mamba decoder, with its more efficient scaling, will experience a much slower degradation. 
This superior scalability and higher initial threshold explicitly demonstrate that the Mamba architecture is a more robust and viable solution for practical, real-time quantum error correction.

\section*{Discussion}
Transformer decoders, while highly accurate, suffer from a computational complexity that scales as $\mathcal{O}(d^4)$ with code distance $d$, creating a significant bottleneck for real-time quantum error correction. In this work, we introduced a neural decoder based on Mamba---a state-space model with $\mathcal{O}(d^2)$ complexity---and showed that it offers a substantially improved balance between speed and accuracy for practical QEC.

In memory experiments using Sycamore hardware data, our Mamba-based neural decoder achieved a logical error rate that matched its Transformer-based counterpart. This finding is crucial, as it demonstrates that the computationally intensive self-attention mechanism can be replaced without compromising---and in fact while preserving---state-of-the-art decoding accuracy in memory experiments. Even before accounting for its speed advantage, the Mamba architecture provides a highly effective foundation for neural decoding.

The Mamba decoder's computational efficiency becomes a decisive advantage in simulated real-time scenarios. In this setting, latency introduces "decoder-induced noise," which severely degrades the Transformer's performance. This penalty, caused by errors accumulating during the Transformer's prolonged processing time, grows rapidly with code distance. The Mamba decoder's speed, however, minimizes this effect. This result underscores that for practical QEC, decoding speed is as fundamental to corrective power as accuracy---a balance the Mamba architecture achieves.

This interplay between speed and performance was further quantified in our error threshold analysis. The Mamba decoder achieved a higher threshold of 0.0104 compared to 0.0097 for the Transformer, indicating that it maintains fault-tolerant scaling under higher physical error rates. A higher threshold directly translates to more lenient hardware noise requirements for achieving fault tolerance. In this sense, the Mamba architecture not only improves decoding speed but also enhances the overall practicality of QEC.

Overall, our results suggest that attention-free architectures represent a promising direction for scalable quantum error correction.
Future work could focus on further optimizing the Mamba architecture—exploring model scaling, advanced training regimes, and hyperparameter tuning—to enhance accuracy without sacrificing speed.
In addition, integrating Mixture-of-Experts (MoE) models~\cite{jacobs1991adaptive, fedus2022switch, zhou2022mixture} presents an intriguing avenue for dynamically allocating computational resources, thereby balancing accuracy and efficiency adaptively.
Beyond the surface code, the same principles can be extended to other stabilizer codes, including color codes~\cite{maskara2019advantages, baireuther2019neural} and Quantum Low-Density Parity-Check (QLDPC) codes~\cite{maan2025machine, blue2025machine} such as Bivariate Bicycle (BB) codes, which are actively investigated as high-rate alternatives but present distinct decoding challenges.
Ultimately, applying these efficient and scalable decoders to complex subroutines like magic state distillation~\cite{sales2025experimental} will be a critical step toward realizing universal fault-tolerant quantum computation.

\section*{Methods}
\label{sec:method}
\subsection*{Model Architecture}

\begin{figure}[t]
    \centering 
    \includegraphics[width=\textwidth]{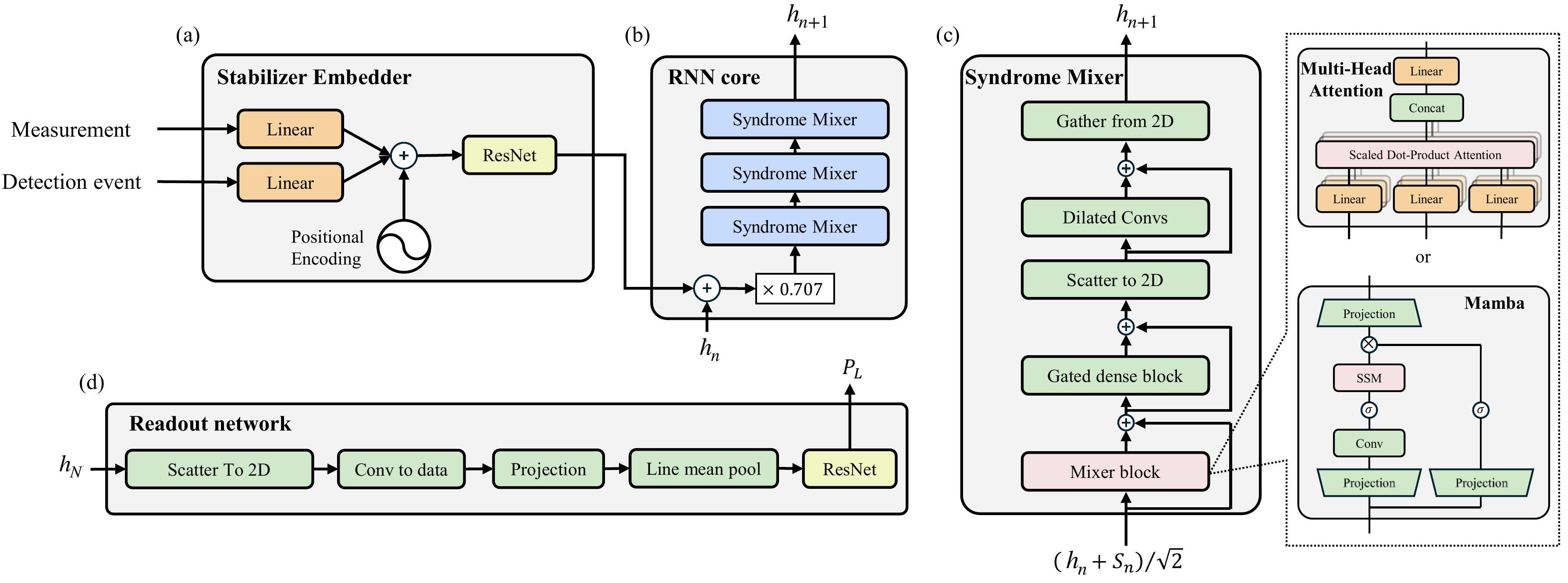}
  \caption{\textbf{(a) Stabilizer Embedder:} Raw stabilizer measurements and detection events are passed through linear layers and combined with positional encodings before being processed by a ResNet to create an embedding $S_n$ for each cycle. \textbf{(b) RNN Core:} The core processes the sequence of embeddings recurrently. At each step, the previous decoder's hidden state $h_n$ is combined with the current syndrome embedding $S_n$ and passed through three Syndrome Mixer layers to produce the updated state $h_{n+1}$. A scaled skip connection is used. \textbf{(c) Syndrome Mixer:} This is the central processing unit. The input is first processed by a Mixer Block, which can be either a Multi-Head Attention module (for the Transformer architecture) or a Mamba module. The output is then passed through a gated dense block and reshaped into a 2D grid to be processed by dilated convolutions, capturing spatial correlations. \textbf{(d) Readout Network:} The final decoder's hidden state $h_N$ is processed by a series of convolutional and dense layers to produce the final prediction for the logical error probability, $P_L$.}
    \label{fig:model_architecture}
\end{figure}

Our decoder follows the recurrent-transformer design of AlphaQubit (see Figure~\ref{fig:fig1}). 
At each error-correction cycle $n$, for each stabilizer $i$, the raw stabilizer measurements and detection events (and optionally analogue I/Q readouts or leakage flags) are embedded into a $d_\text{model}$-dimensional feature $s_{n, i}$ by a stabilizer embedding network. 
The set of embeddings $S_{n}=\{\mathbf{s}_{n,i}\}_{i=1}^{l}$, together with the previous decoder's hidden state ${h}_{n}$, are processed through $L$ Syndrome Mixer layers, producing updated states $h_{n+1}$. 

The Mixer block is a core module designed to learn global dependencies between the induced stabilizer embedding $S_{n}$ and the previous decoder's hidden state $h_n$. 
For our Transformer decoder, this block uses a Multi-Head Attention (MHA) mechanism.
In MHA mechanism, three linear projections are applied to the input vectors to generate Queries, Keys and Values. These are split into $H$ heads and processed via scaled dot-product attention. 
The outputs of each head are concatenated and passed through a final linear transformation. An attention mask may optionally be applied to preserve temporal order or physical locality.

For our Mamba decoder, the Mixer Block is replaced by a Mamba module which performs channel-wise, content-aware mixing. 
It first expands the input dimension and then branches into two parallel paths. 
The first path consists of a standard multi-layer perceptron (MLP) with a linear transformation, followed by Sigmoid Linear Unit activation and another linear transformation, which extracts local features. 
The second path employs a Structured State Space Model (SSM) layer to perform a sequential scan, accumulating long-range contextual information at linear cost. 
The two path outputs are then summed element-wise, projected back to the original dimension.

The global dependency information produced by either the MHA or the Mamba module is then forwarded to a gated dense block, after which they are reshaped onto the $(d + 1) \times (d+1)$ surface code grid and processed with dilated convolutions to model spatial correlations at multiple scales. 
Finally, a readout network predicts logical errors from the final decoder's hidden state. The detailed diagram of neural decoders are shown in Figure~\ref{fig:model_architecture}.

\subsection*{Model Hyperparamters}
Both decoders use a recurrent core consisting of 3 sequential layers, and dropout is not used in any of the models. 
The specific hyperparameters for both decoder architectures are detailed in Table~\ref{tab:tab1}. The descriptions for these parameters are as follows:

\begin{description}
    \item $d_\text{model}$: The dimension of the decoder's hidden state.
    \item $H$: The number of heads in the Multi-Head Attention block.
    \item $d_b$: The dimension of each attention block.
    \item $d_{\text{attn}}$: The total dimension of the attention mechanism.
    \item $d_{\text{mid}}$: The intermediate dimension of the feed-forward network within the Transformer block.
    \item $d_{\text{state}}$: The latent state dimension of the Mamba state-space model.
    \item $d_{\text{conv}}$: The kernel size of the 1D convolution within the Mamba block.
    \item $w_\text{mamba}$: The expansion factor for the internal dimension of the Mamba block.
    \item $L_\text{stab}$: The number of ResNet layers in the Stabilizer Embedder.
    \item $L_\text{read}$: The number ResNet layers in the Readout Nework.
    \item $d_\text{read}$: The hidden dimension of the Readout Network.
    \item $w_\text{gate}$: The widening factor of the feedforward network for the gated dense layers.
    \item $D_\text{conv}$: The list of dilation rates for the 3 dilated convolutional layers.
\end{description}

\begin{table}[h]
    \centering
    \begin{subtable}[t]{.48\textwidth}
        \centering
        \begin{tabular}{ccc}
            \toprule
            \textbf{Hyperparameter} & \textbf{Sycamore} & \textbf{Real-time} \\
            \midrule
            $d_\text{model}$ & 320 & 256 \\
            $H$ & 4 & 4 \\
            $d_b$ & 48 & 48 \\
            $d_{\text{attn}}$ & 32 & 32 \\
            $d_{\text{mid}}$ & 32 & 32 \\
            \bottomrule
        \end{tabular}
        \caption{Trnasformer-specific Hyperparameters} 
        \label{tab:transformer_specific}
    \end{subtable}
    \hfill % Horizontal space
    \begin{subtable}[t]{.48\textwidth}
        \centering
        \begin{tabular}{ccc}
            \toprule
            \textbf{Hyperparameter} & \textbf{Sycamore} & \textbf{Real-time} \\
            \midrule
            $d_\text{model}$ & 320 & 256 \\
            $d_{\text{state}}$ & 16 & 16 \\
            $d_{\text{conv}}$ & 4 & 4 \\
            $w_\text{mamba}$ & 1 & 1 \\
            & & \\ % Empty row for vertical alignment
            \bottomrule
        \end{tabular}
        \caption{Mamba-specific Hyperparameters} 
        \label{tab:mamba_specific}
    \end{subtable}
    
    \vspace{1.5em} 
    
    \centering
    \begin{tabular}{ccc}
        \toprule
        \multicolumn{3}{c}{\textbf{Shared Hyperparameters}} \\
        \cmidrule(lr){1-3} % Rule spanning the 3 columns
        \textbf{Hyperparameter} & \textbf{Sycamore} & \textbf{Real-time} \\
        \midrule
        $L_\text{stab}$ & 2 & 2 \\
        $L_\text{res}$ & 16 & 16 \\
        $d_\text{read}$ & 64 & 48 \\
        $w_\text{gate}$ & 5 & 5  \\
        $D_\text{conv}$ ($d=3$) & [1,1,1] & [1,1,1]\\
        $D_\text{conv}$ ($d=5$) & [1,1,2] & [1,1,2]\\
        $D_\text{conv}$ ($d=7$) & $\cdot$ & [1,2,4] \\
        
        \bottomrule
    \end{tabular}
    
    \caption{Decoder hyperparameters, separated by model-specific (Top) and shared parameters (Bottom).} % <-- MOVED HERE
    \label{tab:tab1}
\end{table}

\subsection*{Training Details for Sycamore Experiments}
Following the approach of AlphaQubit, we first pretrain our decoder on a large synthetic dataset generated from a detector error model (DEM) whose parameters are derived from a Pauli noise model calibrated to match the noise characteristics of the Sycamore quantum processor. 
The noise parameters are extracted from cross-entropy benchmarking (XEB) data provided by Google Quantum AI, which characterizes single-qubit, two-qubit, measurement, and reset errors observed on the hardware. 
Using this calibrated noise model, we generate up to 100 million synthetic syndrome samples for code distances 3 and 5, across both $X$- and $Z$-basis memory experiments. 
The pretraining dataset includes QEC sequences of varying error correcting cycles, with the number of cycles being the odd numbers from 1 to 25 (i.e., $r \in \{1, 3, ..., 25\}$).

To manage the varying cycle lengths, we implement a curriculum learning strategy. 
Initially, the model is trained on shorter error correction cycles, $r \in \{1, 3, 5, 7, 9\}$. 
Every 150,000 training iterations, we progressively expand the training set to include four additional cycles, eventually covering the full range of 1 to 25 cycles. 
This gradual increase in sequence complexity allows the model to first learn error patterns within shorter cycle sequences before modeling how those error patterns evolve over longer durations, which is critical for accurately modeling multi-cycle syndrome evolution.

During pretraining, we use the Lion optimizer~\cite{chen2023symbolic} with an initial learning rate of $5 \times 10^{-6}$ and weight decay of $1 \times 10^{-5}$. 
The learning rate is annealed using a cosine schedule with $T_{\max} = 30{,}000{,}000 / N_\text{batch}$ and a minimum learning rate of $1 \times 10^{-6}$. 
We apply gradient clipping with a maximum norm of $1$ to stabilize training and maintain an exponential moving average (EMA) of the model weights with a decay rate of $0.9999$. 
The EMA model is used during validation and inference, as it provides more stable predictions and better generalization performance.
Training was conducted for up to 50,000 iterations with a batch size of 256 for distance 3, and up to 1,000,000 iterations with a batch size of 128 for distance 5.

After pretraining on synthetic data, we fine-tune the decoder on the actual Sycamore experimental dataset, which provides 50,000 experimental samples for each odd-numbered cycle.
We partition the dataset into a training set (50\% of the data) and a held-out evaluation set (the remaining 50\%). 
The pretrained model weights are loaded as initialization, but we reset the EMA tracker and optimizer state to allow the model to adapt more flexibly to the real hardware data.
Fine-tuning is conducted for 10 epochs using a reduced learning rate of $2 \times 10^{-6}$ to avoid destabilizing the pretrained representations.
We increase the weight decay to $7 \times 10^{-5}$ to provide stronger regularization, which helps prevent overfitting to the smaller real-data training set.

\subsection*{Training Details for Real-Time Decoding Experiments}

For the real-time decoding experiments, we train decoders from scratch using synthetic syndrome data generated according to the SI1000 noise model.
This data is efficiently generated on-the-fly during training using the Stim quantum circuit simulator.

We train separate models for code distances $d \in \{3,5,7\}$ and a physical error rate of $p = 0.002$.
The training sequences consist of $2d+1$ QEC cycles. For evaluation, the logical error rate (LER) is computed over a much longer duration of $8d+4$ cycles. 
To simulate the effect of decoder latency, decoder-induced noise is injected every $2d+1$ cycles throughout the evaluation period, resulting in four distinct noise injections.
We train for 500,000 iterations with a batch size of 256 using the Lion optimizer with an initial learning rate of $5 \times 10^{-6}$ and weight decay of $1 \times 10^{-5}$. 
The learning rate follows a cosine annealing schedule with $T_{\text{max}}=128,000,000/N_\text{batch}$  and a minimum learning rate of $1 \times 10^{-6}$.

To accurately characterize the error threshold of each decoder architecture, we fine-tune the models on multiple physical error rates. 
Starting from the baseline model trained at $p = 0.002$, we fine-tune separate instances for each higher noise level $p \in \{0.006, 0.008, 0.010, 0.012 \}$. 
Fine-tuning is conducted for 250,000 iterations (half the duration of the initial training) with a reduced learning rate of $2 \times 10^{-6}$ to preserve the useful representations learned during pretraining while adapting to the new noise regime.
The cosine annealing schedule is adjusted accordingly with $T_{\max} = 64{,}000{,}000 / N_\text{batch}$. 
This transfer learning approach significantly reduces the computational cost of training models at multiple noise levels while maintaining high performance.

\section*{Acknowledgments}
This work is supported by Institute of Information \& communications Technology Planning \& evaluation (IITP) grant funded by the Korea government (No. 2019-0-00003, Research and Development of Core Technologies for Programming, Running, Implementing and Validating of Fault-Tolerant Quantum Computing System), the National Research Foundation of Korea (RS-2025-02309510), the Ministry of Trade, Industry, and Energy (MOTIE), Korea, under the Industrial Innovation Infrastructure Development Project (RS-2024-00466693), the Basic Science Research Program through the National Research Foundation of Korea (NRF) funded by the Ministry of Education (RS-2025-25419259), and the NVIDIA Academic Grant Program.

\bibliography{reference}
\bibliographystyle{unsrt}
\end{document}